\documentclass{article}
\usepackage{ijcai22}

\usepackage{times}

\usepackage{soul}
\usepackage{url}
\usepackage[hidelinks]{hyperref}
\usepackage[utf8]{inputenc}
\usepackage[small]{caption}
\usepackage{graphicx}
\usepackage{amsmath}
\usepackage{booktabs}
\usepackage{algorithm}
\usepackage{algorithmic}
\usepackage{subfig}
\usepackage{amssymb}
\usepackage{xcolor}

\title{Sparsity Regularization for Cold-Start Recommendation}
\author{
Aksheshkumar Ajaykumar Shah$^1$\footnote{Contact Author}\and
Hemanth Venkateswara$^2$
\affiliations
Center for Cognitive Ubiquitous Computing (CUbiC), Arizona State University\\
\emails
\{aashah16, hemanthv\}@asu.edu
}

\begin{document}

\maketitle

\begin{abstract}
Recently, Generative Adversarial Networks (GANs) have been applied to the problem of Cold-Start Recommendation, but the training performance of these models is hampered by the extreme sparsity in warm user purchase behavior. In this paper we introduce a novel representation for user-vectors by combining user demographics and user preferences, making the model a hybrid system which uses Collaborative Filtering and Content Based Recommendation. Our system models user purchase behavior using  weighted user-product preferences (explicit feedback) rather than binary user-product interactions (implicit feedback). Using this we develop a novel sparse adversarial model, SRLGAN, for Cold-Start Recommendation leveraging the sparse user-purchase behavior which ensures training stability and avoids over-fitting on warm users. We evaluate the SRLGAN on two popular datasets and demonstrate state-of-the-art results. 
\end{abstract}

\section{Introduction}
Recommendation systems identify a fraction of items from a very large inventory of items and recommend them to a user for purchase. 
This task is achieved with the help of the user's past purchase behavior~\cite{BOBADILLA2013109,10.1145/3123266.3123316}. 
Recommendation systems help in enhancing the overall purchase experience for the users.
Recommendation systems can be broadly classified into three categories, viz., Collaborative Filtering-based, Content-based and Hybrid Systems~\cite{Co-Clustering}. 
Among these, collaborative filtering is the most popular approach~\cite{1423975}. 
In collaborative filtering, the past purchase behavior of the user is used for making recommendations~\cite{7927889}. 
The user purchase behavior consists of  user's feedback which can be either implicit or explicit. 
Implicit feedback is binary and represents the user's interaction with the items (purchased or not-purchased)~\cite{2021}. 
Explicit feedback is generally the discrete ratings assigned by the user ranging between 1-to-$C$ for every item purchased. \\

Collaborative filtering can be performed using the matrix factorization~\cite{CFWGAN}, where the user and item information is projected into a $K$-dimensional latent space and their interaction is modelled by the inner product of the latent vectors. 
Matrix factorization attempts to model the linear relationship between users and items and thereby predicts the items a new user would purchase. 
Deep learning based collaborative filtering can model highly nonlinear relationships between users~\cite{lecunn} and items and therefore exhibits superior performance to traditional matrix factorization approaches \cite{NCF,10.1145/2835776.2835837}. \\

Collaborative filtering is successful in recommending items for \emph{warm users} - users whose previous purchase behavior is known. 
However, the approach fails in recommending items for \emph{cold users} - users whose purchase behavior is unknown~\cite{10.1145/3123266.3123316}. 
This problem is popularly known as Cold-Start Recommendation~\cite{10.1145/2484028.2484035}. 
It is of two types, (i) User Cold-Start, and (ii) Item Cold-Start. 
In the former, the user is new and no purchase history is available and in later, the item is new and no user interactions are available. 
When both are compared, the User Cold-Start problem is more complex and also more popular.\\

The Cold-Start Recommendation problem can be tackled by using cross-domain information~\cite{fer}, personal information~\cite{10.1007/s11257-016-9172-z} and social-network information~\cite{Sedhain_Menon_Sanner_Xie_Braziunas_2017} of the user. 
These kinds of models are called as Content-based systems. 
These systems have a limitation where they tend to suggest the same items to all users who are similar even if the users have rated the items differently \cite{lee2019melu}. 
To overcome this problem hybrid systems based on Collaborative Filtering which use content information are widely used~\cite{10.1145/2988450.2988454,10.1145/2792838.2800175}.\\ 

The popular Generative Adversarial Networks (GANs) models from computer vision and natural language processing have been successfully applied towards collaborative filtering~\cite{GANs}. 
The potential of GANs for Collaborative Filtering based Recommendation systems has been demonstrated by methods such as IRGAN \cite{IRGAN}, GraphGan \cite{GraphGAN} and  CFGAN~\cite{CFGAN}. 
In these models the Generator is used to generate the user purchase behavior and the Discriminator is trained to distinguish between the ground truth purchase behavior and the generated purchase behavior (output of the Generator). \\

In this paper we use GANs to implement user-based Cold Start Recommendation. 
We first introduce a novel technique to represent user information by combining the users' demographic information and their preferences. 
This is implemented by creating a Term Frequency - Inverse Document Frequency (TF-IDF) vector for each user. 
The TF-IDF vectors give a weighted score to user attributes including genre preferences which allows us to leverage the advantages of collaborative filtering as well as Content-based methods. 
We model user-purchase behavior using explicit feedback from the users where user-ratings for items are taken into consideration instead of binary implicit feedback which merely indicates if a user has purchased the item or not. 
The Generated user-purchase behavior is in the form of vectors instead of discrete item ID's to take full advantage of adversarial training as also implemented in \cite{CFGAN}. 
In addition we propose a novel sparsity regularization that models the high sparsity in the user-purchase behavior, which in turn ensures stability when training the GAN and avoids the problem of over-fitting on warm users. 
The proposed model is tested on two popular datasets namely MovieLens 100K  and MovieLens 1M~\cite{10.1145/2827872} and demonstrates state-of-the-art results.\\

In summary, the main contributions of this paper are listed as follows:
\begin{itemize}
    \item TF-IDF based representation of user-vectors containing demographic information and preferences.
    \item A novel hybrid Cold-Start Recommendation system combining Collaborative Filtering and Content-based approaches.
    \item A novel sparsity regularization to model the sparsity in the data and ensure training stability. 
    \item Competitive performance on the popular MovieLens 100K and MovieLens 1M datasets. 
\end{itemize}

\section{Related Work}
In this section we discuss Collaborative filtering methods based on deep neural networks and specifically GANs from literature. 
\subsection{Deep Neural Network based Collaborative Filtering Methods}
According to~\cite{Zeroshot}, Cold-Start Recommendation models can be classified into three categories, viz., (i) Similarity-based models, (ii) Matrix Factorization models, and (iii) Feature Mapping models. 
The most popular approach is Matrix factorization, where the user-item relationship matrix is typically factorized into two latent representations which are used to make predictions. 
Matrix factorization models the linear interactions between the latent features of the users and the items. 
Popular methods include, BPR~\cite{rendle2012bpr} and FISM~\cite{10.1145/2487575.2487589}.\\

Deep Neural Network based Collaborative filtering models have been gaining popularity owing to their capabilities of exploiting the non-linear interactions between the features and developing an arbitrary continuous function~\cite{lecunn}. 
Some of the popular top-n recommendation models are CDAE~\cite{10.1145/2835776.2835837} and NCF~\cite{NCF}.\\
\begin{figure*}[t]
    \centering
    \includegraphics[width =\textwidth,height = \textwidth, keepaspectratio]{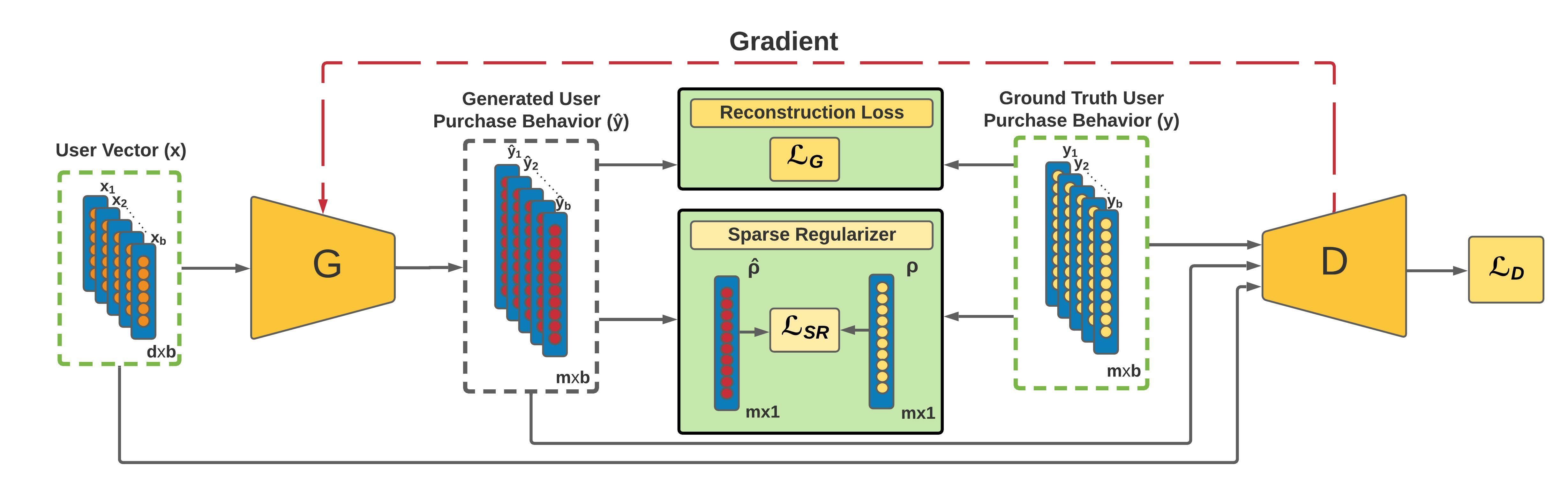}
    \caption{System diagram of the SRLGAN. The input to the Generator module are user attributes of dimensions $(d \times b)$ and the output of the Generator are the predicted user purchase behaviors of dimensions $(m \times b)$. The objective term $\mathcal{L}_G$ (Eq.~\ref{Eq:LSLoss}), is a reconstruction loss to train the Generator to predict user purchase behavior similar to the ground truth. The objective term $\mathcal{L}_D$ is the least squares GAN objective in Eq.~\ref{Eq:LSGAN}. The novel Sparse Regularizer models the sparsity in the user purchase behavior. 
    It aligns the distributions of the predicted and ground truth user purchase behavior with $\mathcal{L}_{SR}$ (Eq.~\ref{Eq:LSR}), by minimizing the KL-divergence between the average generated (predicted) user purchase prediction $\hat{\rho}$ and the average ground truth user purchase prediction $\rho$.}
    \label{fig:SRLGAN}
\end{figure*}

\subsection{Generative Adversarial Networks based Collaborative Filtering Methods}

In domain of recommendation systems Generative Adversarial Networks have also been implemented by some methods for Collaborative filtering. 
They have adopted the adversarial training approach rather than the traditional approach of optimizing the pointwise or pairwise objective function for Collaborative filtering. 
In IRGAN~\cite{IRGAN} and GraphGAN~\cite{GraphGAN}, the Generator attempts to generate discrete item indices and the Discriminator attempts to distinguish between the synthetically generated item from the ground truth item.
When discrete item indices are generated, the Discriminator has difficulty identifying if the item is relevant or not ~\cite{CFGAN}. 
This degrades the quality of the feedback provided by Discriminator to Generator and fails to avail the advantages of GANs. 
To overcome this problem CFGAN~\cite{CFGAN}, proposed a framework where vectors are generated instead of discrete item ID's to prevent Discriminator's confusion. 
This allows the Discriminator to guide the Generator consistently and produce results closer to the ground truth. 
The VAEGAN~\cite{VAEGAN} proposes a model which uses Variational Bayes along with a Generative Adversarial Networks to generate recommendations on strong generalization and weak generalization.
As the above models do not take into account the sparsity in the user-purchase behavior they are prone to over-fitting and instability during training. 
Our proposed model leverages the sparse user-purchase behavior which ensures training stability and avoids over-fitting on warm users.\\

\section{Proposed Approach}
\subsection{Problem Settings}
The Cold-Start Recommendation problem is of two types, User-based and Item-based. 
In this work we discuss user-based models, but the same principles can be extended to the item-based setting as well. 
The User-based Cold-Start Recommendation consists of warm users (known users i.e., purchase behavior is known) and cold users (unknown users i.e., purchase behavior is unknown). 
The user attributes are vectors consisting of the user's demographic information and preferences represented in the form of Term-Frequency-Inverse-Document-Frequency (TF-IDF) vectors which gives weighted representation of the users interests in genres. 
The set of warm users is represented as $\mathcal{X}^s = \{x_i^s\}_{i=1}^{n_s}$, where $n_s$ is number of warm users and $x\in \mathbb{R}^d$. 
The users purchase from $m$ items and also provide a rating from $1,\ldots,C$ for each of the purchased items. 
A user's purchase behavior is the vector of $m$ ratings where a $0$ indicates the item was not purchased and $1$ is the least rating and $C$ is the best rating. 
The ground truth purchase behaviors corresponding to the warm users are $\mathcal{Y}^s = \{y_i^s\}_{i=1}^{n_s}$, where $y \in [0,1]^m$ is a discrete vector of $m$ dimensions (items) where the ratings have been normalized by dividing with $C$. 
The cold users are $\mathcal{X}^u=\{x_i^u\}_{i=1}^{n_u}$, where $n_u$ is the number of cold users. 
Let the space of user attributes be $\mathcal{X}$ and the space of user purchase behaviors be $\mathcal{Y}$. 
The goal of Cold-Start recommendation is to learn a function $\mathcal{F}: \mathcal{X} \rightarrow \mathcal{Y}$, and thereby predict the purchase behavior $\{\hat{y}_i^u\}_{i=1}^{n_u}$ of cold users in $\mathcal{X}^u$. \\

\subsection{Adversarial User Purchase Behavior Generation} 
We propose a generative model to hallucinate the purchase behavior of the users. 
We apply a conditional Generative Adversarial Network (GAN) which takes user attributes as input to the Generator and outputs user purchase behavior corresponding to the user. 
The Discriminator is trained to distinguish ground-truth user purchase behavior vs. generated user behavior. 
The components of the GAN model are described below. 
\subsubsection{User Purchase Behavior Generator}
The conditional generator is a mapping $G_{\theta_g}: \mathcal{X} \rightarrow \mathcal{Y}$ with parameters $\theta_g$ where, $\mathcal{X}$ and $\mathcal{Y}$ are the space of user attributes and user purchase behavior, respectively. 
The Generator is trained to predict user purchase behavior for warm users ($\hat{y}^s \leftarrow G_{\theta_g}(x^s)$) as well as for cold users ($\hat{y}^u \leftarrow G_{\theta_g}(x^u)$). 
The ground truth warm user purchase behavior $y^s$ can be used to supervise the Generator with a least squares objective, 
\begin{equation}
    \mathcal{L}_{G} = \min _{\theta_g} \frac{1}{n_s} \sum_{i=1}^{n_s}(y_i^s - G_{\theta_g}(x_i^s))^2.
    \label{Eq:LSLoss}
\end{equation} 


\subsubsection{User Purchase Behavior Discriminator} 

We train the conditional GAN with an adversarial Discriminator which is inspired by LSGAN to ensure training stability \cite{mao2017squares}. 
The loss function for this GAN is a least squares error in place of the standard binary cross entropy loss as it gives a better gradient for the model to learn. 
The input to the Discriminator is a user attribute vector ($x$) concatenated with the user purchase vector ($y$). 
The Discriminator learns a mapping $D_{\theta_d}: \mathcal{X}\times\mathcal{Y} \rightarrow [0,1]$, where $\theta_d$ are the parameters of the Discriminator. 
The Discriminator is trained to distinguish between the ground truth user purchase behavior $(y)$ and the generated user purchase behavior $(\hat{y})$. 
The objective function for training  Discriminator and Generator is, 
\begin{multline}
    \mathcal{L}_D = \min_{\theta_g}\max_{\theta_d}\frac{1}{2}\mathbb{E}_{(x,y)\sim(\mathcal{X}^s,\mathcal{Y}^s)} \big[(D_{\theta_d}(x,y)-1)^2\big] \\+\frac{1}{2}\mathbb{E}_{x\sim\mathcal{X}^s}\big[(D_{\theta_d}(x,G_{\theta_g}(x))^2\big].
    \label{Eq:LSGAN}
\end{multline}
The objective function is maximized w.r.t. the Discriminator's parameters, $\theta_d$ and minimized w.r.t. the Generator's parameters, $\theta_g$. 

\subsection{Sparsity Regularization}
When there are a large number of items, the purchase behavior of a random user is usually highly sparse with lots of $0$s in the purchase behavior vector, i.e., the user has purchased only a few items. 
The user purchase behavior is a discrete vector $y\in[0,1]^m$ with high sparsity (multiple zero values or multiple not purchased items). 
We observed instability in training with poor convergence when we attempted to train the Cold-Start recommendation GAN without taking into account sparsity.   
In addition, GAN based models are biased towards warm users leading to poor generalization and can often end up in a mode collapse where the Generator outputs the same user-purchase behavior for all users. 
With inspiration from the sparse autoencoder \cite{ng2011sparse}, we introduce a novel regularization to model the sparsity in user purchase behavior by aligning the distributions of the predicted purchase behavior and the ground truth purchase behavior. 
We treat the purchase behavior for each item as a Bernoulli random variable with an unknown mean. 
We estimate the mean purchase behavior for all the items from the training set for $n_s$ samples $\{y^s_i\}_{i=1}^{n_s}$. 
The average purchase behavior for the dataset is $\rho$, where, 
\begin{equation}
\rho = \frac{1}{n_s}\sum_{i=1}^{n_s} y^s_i.
\label{Eq:rho}
\end{equation}
$\rho = [\rho^{(1)},\rho^{(2)}, \ldots, \rho^{(m)}]^\top$ is the average ground truth purchase behavior where $\rho^{(i)} \in [0,1]$ is the average purchase behavior for item $i$. 
Likewise, the average predicted purchase behavior for a batch size $b$ is given by $\hat{\rho} = \frac{1}{b}\sum_{i=1}^b \hat{y}_i$. 
This is estimated from the output of the Generator. 
The Kullback-Leibler divergence $\text{KL}(\rho^{(i)}||\hat{\rho}^{(i)})$, is a measure of divergence between a Bernoulli random variable with mean $\rho^{(i)}$ and a Bernoulli random variable with mean $\hat{\rho}^{(i)}$. 
We propose a sparsity regularization to align the distributions of the predicted user purchase behaviors with the ground truth user purchase behaviors, which we model as Bernoulli random variables. 
This regularize the user purchase prediction by minimizing the divergence between the distributions of the predicted and ground truth user purchase behaviors. 
The sparsity regularization for a batch of $m$ user purchase predictions is defined as, 
\begin{flalign}
\mathcal{L}_{SR} &= \sum_{i=1}^m\text{KL}(\rho^{(i)}||\hat{\rho}^{(i)})\notag\\
 &=\sum_{i=1}^m\rho^{(i)}\log\frac{\rho^{(i)}}{\hat{\rho}^{(i)}} + (1-\rho^{(i)})\log\frac{(1-\rho^{(i)})}{(1-\hat{\rho}^{(i)})}.
\label{Eq:LSR}
\end{flalign}
The divergence $\text{KL}(\rho^{(i)}||\hat{\rho}^{(i)}) = 0$ if $\rho^{(i)} = \hat{\rho}^{(i)}$, otherwise it increases monotonically to infinity as the value of $\hat{\rho}^{(i)}$ diverges from $\rho^{(i)}$. 
The regularization prevents the Generator from getting over-fitted on the warm users and also helps in avoiding the problem of mode collapse which are two of the prominent problems in the training of GAN based recommendation systems. 

\subsection{SRLGAN Objective Function}
We implement a user-based Cold-Start Recommendation model called the \textbf{S}parse \textbf{R}egu\textbf{L}arized \textbf{G}enerative \textbf{A}dversarial \textbf{N}etwork (SRLGAN). 
The SRLGAN is trained using content-based user representation captured in the form of user preferences and the Collaborative filtering gathered from the user's purchase behavior. 
The training of the SRLGAN is driven by 3 objective functions. 
The Generator component in the SRLGAN is guided by the least squares loss to predict the ground truth user purchase behavior for warm users (see Eq. ~\ref{Eq:LSLoss}). 
The Discriminator and the Generator are trained using the adversarial Least Squares objective (see Eq.~\ref{Eq:LSGAN}). 
The SRLGAN leverages the sparsity in the user purchase behaviour distribution to train the Generator and avoid over fitting to warm users (see Eq.~\ref{Eq:LSR}).
The overall objective function of the SRLGAN is, 
\begin{equation}
    \mathcal{L}_{G} + \mathcal{L}_D + \beta\mathcal{L}_{SR},
    \label{Eq:SRLGAN}
\end{equation}
where, $\beta\geq0$ is a hyper-parameter that controls the importance of the sparse regularization. 
Larger the value of $\beta$, higher the sparsity in the predicted user purchase behavior. 
The value of $\beta$ is estimated using cross validation when predicting the purchase of warm users. 
The SRLGAN is depicted in Fig.~\ref{fig:SRLGAN}. 

\begin{table*}[t]
    \centering
    \resizebox{\linewidth}{!}{
    \begin{tabular}{|l|c|c|c|c|c|c|c|c|}
    \hline
     & \multicolumn{3}{|c|}{\textbf{MovieLens 100K}} &\multicolumn{3}{|c|}{\textbf{MovieLens 1M}}\\ \hline 
    &\small\small P@5\space\space \small\small P@20 &\small\small N@5\space\space \small\small N@20 &\small\small M@5\space\space \small\small M@20 &\small\small P@5\space\space \small\small P@20 &\small\small N@5\space\space \small\small N@20 &\small\small M@5\space\space \small\small M@20\\ \hline
     \small ItemPop &\small.181\space\space\space\space  \small.138 &\small.163\space\space\space\space  \small.195 &\small.254\space\space\space\space  \small.292  &\small.157\space\space\space\space  \small.121 &\small.154\space\space\space\space  \small.181 &\small.252\space\space\space\space  \small.297\\ \hline
    \small BPR~\cite{rendle2012bpr} &\small.348\space\space\space\space  \small.236 &\small.370\space\space\space\space  \small.380 &\small.556\space\space\space\space  \small.574  &\small.341\space\space\space\space  \small.252 &\small.349\space\space\space\space  \small.362 &\small.537\space\space\space\space  \small.556 \\\hline
    \small FISM~\cite{10.1145/2487575.2487589} &\small.426\space\space\space\space  \small.285 &\small.462\space\space\space\space  \small.429 &\small.674\space\space\space\space  \small.685  &\small.420\space\space\space\space  \small.302 &\small.443\space\space\space\space  \small.399 &\small.637\space\space\space\space  \small.651\\  \hline
    \small CDAE~\cite{10.1145/2835776.2835837} &\small.433\space\space\space\space  \small.287 &\small.465\space\space\space\space  \small.425 &\small.664\space\space\space\space  \small.674  &\small.419\space\space\space\space  \small.307 &\small.439\space\space\space\space  \small.401 &\small.629\space\space\space\space  \small.644\\ \hline
    \small GraphGAN~\cite{GraphGAN} &\small.212\space\space\space\space  \small.151 &\small.183\space\space\space\space  \small.249 &\small.282\space\space\space\space  \small.312  &\small.178\space\space\space\space  \small.194 &\small.205\space\space\space\space  \small.184 &\small.281\space\space\space\space  \small.316\\ \hline
    \small IRGAN~\cite{IRGAN} &\small.312\space\space\space\space  \small.221 &\small.342\space\space\space\space  \small.368 &\small.536\space\space\space\space  \small.523  &\small.263\space\space\space\space  \small.214 &\small.264\space\space\space\space  \small.246 &\small.301\space\space\space\space  \small.338\\ \hline
    \small CFGAN~\cite{CFGAN} &\small.444\space\space\space\space  \small.292 &\small.476\space\space\space\space  \small.433 &\small.681\space\space\space\space  \small.693  &\small.432\space\space\space\space  \small.309 &\small.455\space\space\space\space  \small.406 &\small.647\space\space\space\space  \small.660\\ \hline
    \small VAEGAN~\cite{VAEGAN} &\small-\space\space\space\space  \small- &\small.468\space\space\space\space  \small.437 &\small.688\space\space\space\space  \textbf{\small.700}  &\small- \space\space\space\space  \small- &\small.465\space\space\space\space  \small.416 &\small.663\space\space\space\space  \small.676\\ \hline
    \small SRLGAN [Ours]&\textbf{\small.521}\space\space\space\space  \textbf{\small.444} &\textbf{\small.530}\space\space\space\space  \textbf{\small.466} &\textbf{\small.690}\space\space\space\space  \small.699  &\textbf{\small.499}\space\space\space\space  \textbf{\small.436} &\textbf{\small.504}\space\space\space\space  \textbf{\small.453} &\textbf{\small.674}\space\space\space\space  \textbf{\small.683}\\ \hline
    \end{tabular}}
    \caption{Comparison of the performance of SRLGAN on MovieLens 100K and MovieLens 1M datasets with respect to the baselines. Here, P@n stands for Precision (P), N@n stands for Normalized Discounted Cumulative Gain (N) and M@n stands for Mean Reciprocal Rank (M) when considering n items. The value of n = 5,20. The best results are highlighted in bold.}
    \label{tab:results}
\end{table*}
\section{Experiments and Analysis}
\subsection{Datasets}
\begin{table}[ht]
    \centering
    \begin{tabular}{|l|c|c|c|}
    \hline
    \textbf{Dataset}     & \textbf{Users} & \textbf{Items} & \textbf{\% Sparsity} \\ \hline
     MovieLens 100K    & 943 & 1682 & 93.69\\ \hline
     MovieLens 1M & 6040 & 3952 & 95.80 \\ \hline
    \end{tabular}
    \caption{Details of the Datasets. The \% Sparsity is the percentage of $0$s in the user purchase behavior space $\mathcal{Y}$.}
    \label{tab:dataset}
\end{table}
The SRLGAN is evaluated using two popular datasets, MovieLens 100k and MovieLens 1M~\cite{10.1145/2827872}. 
In our experiment we divide the datasets into user attributes space, $\mathcal{X}$ and the user purchase behavior space, $\mathcal{Y}$. 
The user attributes space $\mathcal{X}$ consists of user vectors $x$ of dimension 103 and 48 for MovieLens 100K and MovieLens 1M, respectively.
The user vector consists of user attributes such as demographic information of the users including ages, gender, occupation and 19 movie genres as defined in the dataset. 
Since all the details provided related to the user are not equally important, the user vector is represented in the form of TF-IDF vectors. 
This helps in giving higher weighted score to the attributes that are dominant.
Movie tags of users' previously seen movies are used to indicate the presence of each genre.
Term frequency is the count of a user attribute. 
In order to reduce the importance of common attributes and highlight the dominant attributes, inverse document frequency is calculated using the frequency of attributes across all users.
The final user attribute vector is an element-wise product of the term frequency vector and the inverse document frequency vector.
The user purchase behavior space consists of the ratings given by the user to the products. 
We have considered explicit representation of user interaction as it gives us graded score of users' interest in the product. 
The details related to number of unique users and items in the dataset is defined in Table~\ref{tab:dataset}. 
To ensure a fair comparison across different approaches, 20\% of the users were selected randomly as cold users (test set) and the remaining 80\% were treated as warm users. 
In real world scenarios, the user vectors for cold users can be estimated by asking the users to note their preferences for different genres.  

\subsection{Implementation Details and Performance Metrics}
\subsubsection{Evaluation Metrics}
We have adopted three common metrics which are used for the evaluation of Top-n recommendation systems \cite{CFGAN}, \cite{VAEGAN}. 
(P@n) is the Precision P when considering n items. 
(N@n) is the Normalized Discounted Cumulative Gain (NDCG) N for n items and (M@n) is the Mean Reciprocal Rank M for n items. 
The last two metrics are based on the rank of the correctly predicted item, where lower the rank, higher is the score.  
The metrics are evaluated for n at 5 and 20. \\

\subsubsection{Baselines} 
We compare the performance of the SRLGAN against multiple Top-n recommendation models. 
\textbf{ItemPop} is a non-personalized method in which the items are ranked in descending order of the number of purchase records. 
In the \textbf{BPR}~\cite{rendle2012bpr} method, the relative order of preferences of the purchased and non-purchased item pairs is optimized. 
In the \textbf{FISM}~\cite{10.1145/2487575.2487589} approach, item-item similarity matrix is learned as a product of two low-dimensional latent factor matrices. 
In the \textbf{CDAE}~\cite{10.1145/2835776.2835837} method, collaborative filtering is performed by using a denoising autoencoder along with user-specific latent features. 
\textbf{IRGAN}~\cite{IRGAN} is a popular method that applies GANs for the purpose of collaborative filtering. 
The \textbf{GraphGAN}~\cite{GraphGAN} approach uses graph softmax, which is softmax applied over a graph structure. 
\textbf{CFGAN}~\cite{CFGAN} is a vector-wise adversarial training approach that is closely related to the SRLGAN. 
The \textbf{VAEGAN}~\cite{VAEGAN} applies a variational autoencoder with a contractive loss to generate recommendations. \\

\begin{algorithm}[t]
\caption{Training the SRLGAN}
\label{alg:algorithm}
\textbf{Input}: $x^s \in \mathcal{X}^s$, $y^s \in \mathcal{Y}^s$, $\rho$ \\
\textbf{Constants}: $\beta$, $b$, $n_e$, $n_G$, $n_D$  \\
\textbf{Output}: $G_{\theta_g}$
\begin{algorithmic}[1] 
\STATE Initialize $G_{\theta_g}$and $D_{\theta_d}$
\FOR{$n_e$ iterations} 
\STATE Update $G_{\theta_g}$ using $\{x^s_i,y^s_i\}_{i=1}^b$ with  Eq.~\ref{Eq:LSLoss} 
\ENDFOR
\WHILE{Not Converged}
\FOR{$n_D$ iterations} 
\STATE Generate $\hat{y_i} \leftarrow G_{\theta_g}(x_i^s)$ using $\{x^s_i,y^s_i\}_{i=1}^b$
\STATE Update $D_{\theta_d}$ and $G_{\theta_g}$ using Eq.~\ref{Eq:LSGAN}
\ENDFOR
\FOR{$n_G$ iterations}
\STATE Generate $\hat{y_i} \leftarrow G_{\theta_g}(x_i^s)$ using $\{x^s_i,y^s_i\}_{i=1}^b$
\STATE calculate $\hat{\rho}$ using Eq.~\ref{Eq:rho} \& Update $G_{\theta_g}$ using Eq.~\ref{Eq:SRLGAN}
\ENDFOR
\ENDWHILE
\STATE \textbf{Return} $G_{\theta_g}$
\end{algorithmic}
\end{algorithm} 

\begin{figure*}[ht!]
    \centering
    \subfloat[]{\label{fig:Precision parameter1}
    \includegraphics[trim = {22 0 46 35},clip, width=.235\textwidth]{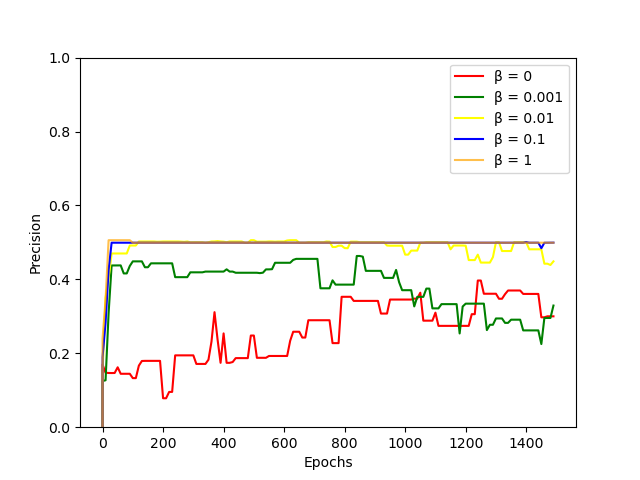}
    }
    \hspace{0.001\textwidth}
    \subfloat[]{\label{fig:Precision parameter2}
    \includegraphics[trim = {22 0 46 35},clip, width=.235\textwidth]{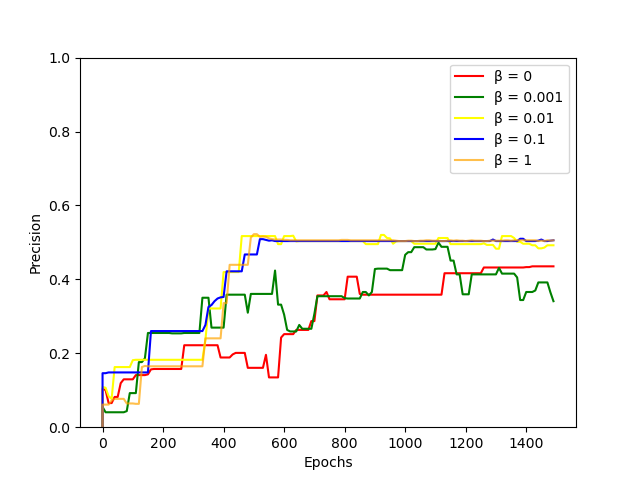}
    }
    \hspace{0.001\textwidth}
    \subfloat[]{\label{fig:Precision parameter3}
    \includegraphics[trim = {22 0 46 35},clip, width=.235\textwidth]{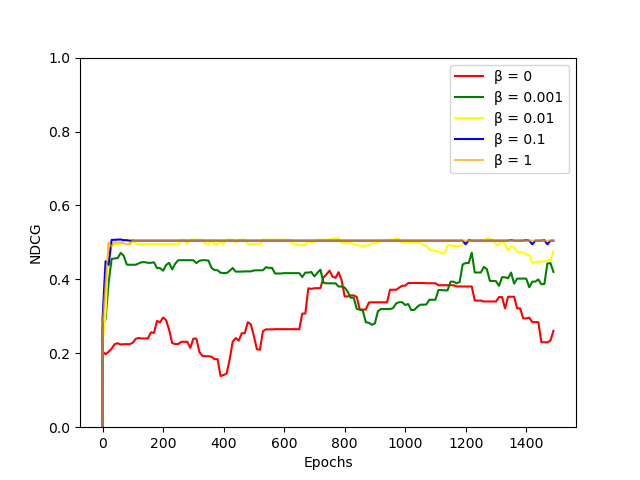}
    }
    \hspace{0.001\textwidth}
    \subfloat[]{\label{fig:Precision parameter4}
    \includegraphics[trim = {22 0 46 35},clip, width=.235\textwidth]{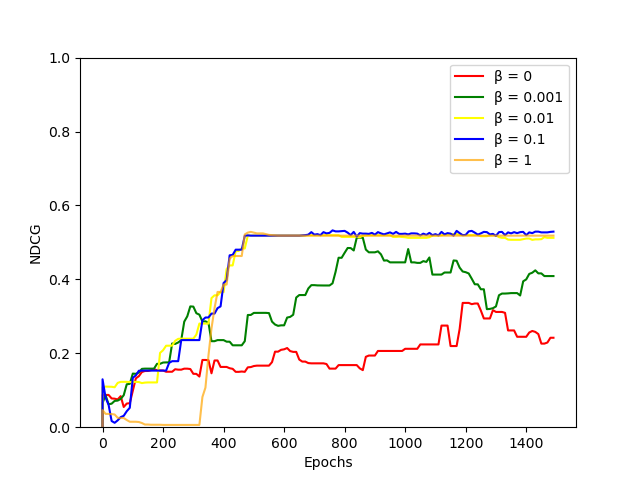}
    }
    \caption{Depicts the training curves for Precision and NDCG as the value of $\beta$ is varied.   Fig.~\ref{fig:Precision parameter1},~\ref{fig:Precision parameter3} are for MovieLens 1M and Fig.~\ref{fig:Precision parameter2},~\ref{fig:Precision parameter4} are for MovieLens 100K.  $\beta=0.1$ produces the best results across both the datasets.}
    \label{fig:Parameter Tuning}
\end{figure*}
\begin{figure*}[!th]
    \centering
    \subfloat[]{\label{fig:Precision}
    \includegraphics[width=.31\textwidth]{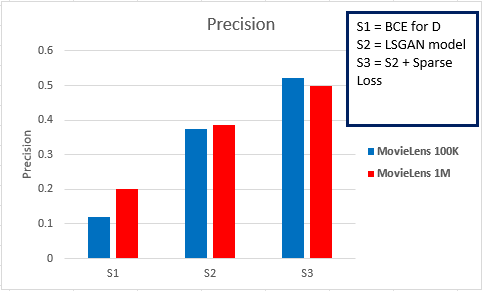}
    }
    \hspace{0.001\textwidth}
    \subfloat[]{\label{fig:NDCG}
    \includegraphics[width=.31\textwidth]{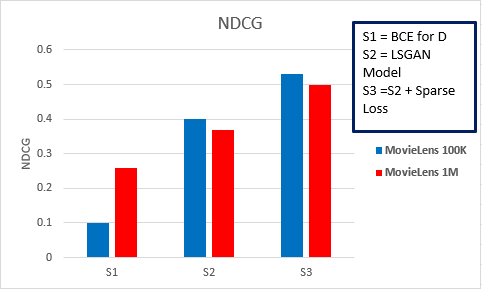}
    }
    \hspace{0.001\textwidth}
    \subfloat[]{\label{fig:MRR}
    \includegraphics[width=.31\textwidth]{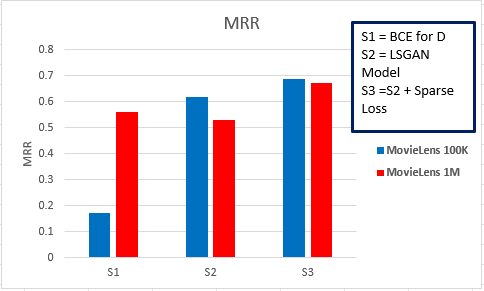}
    }
    \caption{Depicts the ablation results for three different models S1, S2 and S2, each having different components of the SRLGAN. Fig.~\ref{fig:Precision} depicts the variation in Precision, Fig.~\ref{fig:NDCG} showcases variations in NDCG scores, and Fig.~\ref{fig:MRR} displays the variations in the MRR scores across the three models for both the datasets. }
    \label{fig:Ablation Study}
\end{figure*} 

\subsubsection{Model Details}
The Generator and the Discriminator in the SRLGAN are implemented as fully-connected neural network layers with LeakyRelu activations except for the final layers, which have Sigmoid activations. 
The user data is of dimensions $(d\times b)$, where $d$ is the dimension of the user vector and $b$ is the batch size. 
The user-purchase behavior is of dimensions $(m\times b)$, where $m$ is the number of items. 
The number of neurons in the Generator layers are, $[d,512, 1024, 1024, m]$. 
The number of neurons in the Discriminator layers are, $[m+d, 2048, 512, 128, 1]$. 
A dropout of 0.4 is applied to all the hidden layers of the Discriminator to prevent over-fitting, and the learning rate is $10^{-6}$. 
The Generator is initially trained using the $\mathcal{L}_G$ loss to lead the Discriminator. 
This ensures stable training for the SRLGAN. 
Algorithm \ref{alg:algorithm} outlines the training procedure for the SRLGAN. 
For a given user $x$, the output of the Generator $(\hat{y}\leftarrow G_{\theta_g}(x))$ is of dimension $m$, where each component $\{\hat{y}^{(i)}\}_{i=1}^m$ denotes the probability of the user $x$ purchasing item $i$. 
The components of $\hat{y}$ are sorted in descending order to gather the user's preference and to estimate the metrics Precision, NDCG and MRR.

\subsection{Results and Analysis}

\subsubsection{Comparison Against Baselines}
Table \ref{tab:results} outlines the performance of SRLGAN against multiple baselines for the two datasets. 
The SRLGAN outperforms the previous state-of-the-art baselines such as  CFGAN and VAEGAN methods. 
When recommending top-5 items (P@5), the SRLGAN achieves  an improvement of 17.3\% for the Movielens 100K and an improvement of 15.5\% for the MovieLens 1M compared to the best model (CFGAN).  
Similarly, the SRLGAN gains 52\% and 41.1\% for MovieLens 100K and MovieLens 1M, respectively, when predicting the top-20 recommendations (P@20). 
The values in bold in Table \ref{tab:results} demonstrate that the SRLGAN outperforms the other approaches by significant margins across all three metrics for Cold Start recommendation.

\subsubsection{Hyper-parameter Sensitivity}
The $\beta$ hyper-parameter controls the importance of sparsity regularization. 
We estimate the optimal value of $\beta=0.1$ using cross-validation by predicting the purchase behavior of 10\% of warm users in the training set after training with the remaining users.  
Fig.~\ref{fig:Parameter Tuning} plots the evolution of Precision and the NDCG scores for both the datasets for different values of $\beta$. 

\subsubsection{Ablation Study}
We also report an ablation study of the SRLGAN in Fig.~\ref{fig:Ablation Study} to estimate the importance of different components in the SRLGAN. 
We treat the S1 model as the baseline. 
Here the loss function driving the training is the standard binary cross entropy (BCE) loss for the GAN and a reconstruction loss (Eq.~\ref{Eq:LSLoss}) to train the Generator. 
In the S2 model we replace the BCE loss with the least squares loss (Eq.~\ref{Eq:LSGAN}). 
Model S3 is model S2 along with the sparse regularization (Eq.~\ref{Eq:LSR}). 
The plots in Fig.~\ref{fig:Ablation Study} depict a steady improvement in Precision, NDCG and MRR with the introduction of each of the components across both the datasets. 
\section{Conclusions}
In this paper we have proposed a novel Cold-Start Recommendation model, SRLGAN which leverages the sparsity in the user purchase behavior distribution during training. 
The SRLGAN model employs a KL-divergence based sparse penalty which reduces the dissimilarity between the ground truth user purchase behavior distribution and the generated user purchase behavior distribution thereby implementing stable  collaborative filtering for highly sparse datasets. 
Extensive experiments on two popular benchmark datasets demonstrate the SRLGAN outperforms popular approaches. 
It also verifies that our proposed sparse penalty prevents the model from over-fitting and getting into mode collapse. 
In the future we intend to test the performance of the SRLGAN on the related problem of item-based Cold-Start recommendation which is similar to the user-based problem where the item-user interaction data is highly sparse. 

\bibliographystyle{named}
\bibliography{SRLGAN}

\begin{thebibliography}{}

\bibitem[\protect\citeauthoryear{Adomavicius and Tuzhilin}{2005}]{1423975}
Gediminas Adomavicius and Alexander Tuzhilin.
\newblock Toward the {N}ext {G}eneration of {R}ecommender {S}ystems: A {S}urvey
  of the {S}tate-of-the-art and {P}ossible {E}xtensions.
\newblock {\em IEEE Trans. KDE}, 2005.

\bibitem[\protect\citeauthoryear{Bobadilla \bgroup \em et al.\egroup
  }{2013}]{BOBADILLA2013109}
J.~Bobadilla, F.~Ortega, A.~Hernando, and A.~Gutiérrez.
\newblock Recommender systems survey.
\newblock {\em Knowledge-Based Systems}, 46, 2013.

\bibitem[\protect\citeauthoryear{Chae \bgroup \em et al.\egroup }{2018}]{CFGAN}
Dong-Kyu Chae, Jin-Soo Kang, Sang-Wook Kim, and Jung-Tae Lee.
\newblock C{FGAN}: {A} {G}eneric {C}ollaborative {F}iltering {F}ramework based
  on {G}enerative {A}dversarial {N}etworks.
\newblock {\em CIKM}, 2018.

\bibitem[\protect\citeauthoryear{Cheng \bgroup \em et al.\egroup
  }{2016}]{10.1145/2988450.2988454}
Heng-Tze Cheng, Levent Koc, Jeremiah Harmsen, Tal Shaked, Tushar Chandra,
  Hrishi Aradhye, Glen Anderson, Greg Corrado, Wei Chai, Mustafa Ispir, Rohan
  Anil, Zakaria Haque, Lichan Hong, Vihan Jain, Xiaobing Liu, and Hemal Shah.
\newblock Wide \& {D}eep {L}earning for {R}ecommender {S}ystems.
\newblock In {\em Proc. of the 1st Workshop on Deep Learning for Recommender
  Systems}, DLRS 2016. Association for Computing Machinery, 2016.

\bibitem[\protect\citeauthoryear{Feng \bgroup \em et al.\egroup
  }{2019}]{Co-Clustering}
Liang Feng, Qianchuan Zhao, and Cangqi Zhou.
\newblock Improving {P}erformance of {T}op-{N} {R}ecommendations with
  {C}o-{C}lustering method.
\newblock {\em Expert Systems With Applications}, 2019.

\bibitem[\protect\citeauthoryear{Fern\'{a}ndez-Tob\'{\i}as \bgroup \em et
  al.\egroup }{2016}]{10.1007/s11257-016-9172-z}
Ignacio Fern\'{a}ndez-Tob\'{\i}as, Matthias Braunhofer, Mehdi Elahi, Francesco
  Ricci, and Iv\'{a}n Cantador.
\newblock Alleviating the {N}ew {U}ser {P}roblem in {C}ollaborative {F}iltering
  by {E}xploiting {P}ersonality {I}nformation.
\newblock {\em User Modeling and User-Adapted Interaction}, 2016.

\bibitem[\protect\citeauthoryear{Fernández-Tobías \bgroup \em et al.\egroup
  }{2012}]{fer}
Ignacio Fernández-Tobías, Iván Cantador, Marius Kaminskas, and Francesco
  Ricci.
\newblock Cross-domain recommender systems: A survey of the state of the art.
\newblock {\em Proc. of the 2nd Spanish Conf. on Information Retrieval}, 2012.

\bibitem[\protect\citeauthoryear{Goodfellow \bgroup \em et al.\egroup
  }{2014}]{GANs}
Ian~J. Goodfellow, Jean Pouget-Abadie, Mehdi Mirza, Bing Xu, David
  Warde-Farley, Sherjil Ozair, Aaron Courville, and Yoshua Bengio.
\newblock Generative {A}dversarial {N}etworks.
\newblock {\em Neurips}, 2014.

\bibitem[\protect\citeauthoryear{Harper and Konstan}{2015}]{10.1145/2827872}
F.~Maxwell Harper and Joseph~A. Konstan.
\newblock The {M}ovie{L}ens {D}atasets: {H}istory and {C}ontext.
\newblock {\em ACM Trans. Interact. Intell. Syst.}, 2015.

\bibitem[\protect\citeauthoryear{HE \bgroup \em et al.\egroup }{2017}]{NCF}
Xiangnan HE, Lizi Liao, Hanwang Zhang, Liqiang Nie, Xia Hu, and Tat-Seng Chua.
\newblock Neural {C}ollaborative {F}iltering.
\newblock {\em WWW}, 2017.

\bibitem[\protect\citeauthoryear{Kabbur \bgroup \em et al.\egroup
  }{2013}]{10.1145/2487575.2487589}
Santosh Kabbur, Xia Ning, and George Karypis.
\newblock Fism: {F}actored {I}tem {S}imilarity {M}odels for {T}op-{N}
  {R}ecommender {S}ystems.
\newblock In {\em SIGKDD}, 2013.

\bibitem[\protect\citeauthoryear{Kouki \bgroup \em et al.\egroup
  }{2015}]{10.1145/2792838.2800175}
Pigi Kouki, Shobeir Fakhraei, James Foulds, Magdalini Eirinaki, and Lise
  Getoor.
\newblock Hy{PER}: A {F}lexible and {E}xtensible {P}robabilistic {F}ramework
  for {H}ybrid {R}ecommender {S}ystems.
\newblock In {\em Proc. of the 9th ACM Conf. on Recommender Systems}, RecSys
  '15. Association for Computing Machinery, 2015.

\bibitem[\protect\citeauthoryear{LeCun \bgroup \em et al.\egroup
  }{2015}]{lecunn}
Yann LeCun, Yoshua Bengio, and Geoffrey Hinton.
\newblock Deep learning, 2015.

\bibitem[\protect\citeauthoryear{Lee \bgroup \em et al.\egroup
  }{2019}]{lee2019melu}
Hoyeop Lee, Jinbae Im, Seongwon Jang, Hyunsouk Cho, and Sehee Chung.
\newblock Me{LU}: {M}eta-{L}earned {U}ser {P}reference {E}stimator for
  {C}old-{S}tart {R}ecommendation, 2019.

\bibitem[\protect\citeauthoryear{Li \bgroup \em et al.\egroup
  }{2017}]{10.1145/3123266.3123316}
Jingjing Li, Ke~Lu, Zi~Huang, and Heng~Tao Shen.
\newblock Two {B}irds {O}ne {S}tone: {O}n {B}oth {C}old-{S}tart and
  {L}ong-{T}ail {R}ecommendation.
\newblock In {\em Proc. of the 25th ACM Int. Conf. on Multimedia}, 2017.

\bibitem[\protect\citeauthoryear{Li \bgroup \em et al.\egroup
  }{2019}]{Zeroshot}
Jingjing Li, Mengmeng Jing, Ke~Lu, Lei Zhu, Yang Yang, and Zi~Huang.
\newblock From zero-shot learning to cold-start recommendation.
\newblock {\em Proceedings of the {AAAI} Conf. on {AI}}, 2019.

\bibitem[\protect\citeauthoryear{Li \bgroup \em et al.\egroup }{2020}]{CFWGAN}
Rui Li, Fulan Qian, Xiuquan Du, Shu Zhao, and Yanping Zhang.
\newblock A {C}ollaborative {F}iltering {R}ecommendation framework based on
  {W}asserstein {GAN}.
\newblock {\em Journal of Physics: Conference Series}, 2020.

\bibitem[\protect\citeauthoryear{Lin \bgroup \em et al.\egroup
  }{2013}]{10.1145/2484028.2484035}
Jovian Lin, Kazunari Sugiyama, Min-Yen Kan, and Tat-Seng Chua.
\newblock Addressing {C}old-{S}tart in {A}pp {R}ecommendation: {L}atent {U}ser
  {M}odels {C}onstructed from {T}witter {F}ollowers.
\newblock In {\em SIGIR}, 2013.

\bibitem[\protect\citeauthoryear{Mao \bgroup \em et al.\egroup
  }{2017}]{mao2017squares}
Xudong Mao, Qing Li, Haoran Xie, Raymond Y.~K. Lau, Zhen Wang, and Stephen~Paul
  Smolley.
\newblock Least {S}quares {G}enerative {A}dversarial {N}etworks, 2017.

\bibitem[\protect\citeauthoryear{Ng and others}{2011}]{ng2011sparse}
Andrew Ng et~al.
\newblock Sparse autoencoder.
\newblock {\em CS294A Lecture notes}, 72(2011), 2011.

\bibitem[\protect\citeauthoryear{Rendle \bgroup \em et al.\egroup
  }{2012}]{rendle2012bpr}
Steffen Rendle, Christoph Freudenthaler, Zeno Gantner, and Lars Schmidt-Thieme.
\newblock B{PR}: {B}ayesian {P}ersonalized {R}anking from {I}mplicit
  {F}eedback, 2012.

\bibitem[\protect\citeauthoryear{Sedhain \bgroup \em et al.\egroup
  }{2017}]{Sedhain_Menon_Sanner_Xie_Braziunas_2017}
Suvash Sedhain, Aditya Menon, Scott Sanner, Lexing Xie, and Darius Braziunas.
\newblock Low-{R}ank {L}inear {C}old-{S}tart {R}ecommendation from {S}ocial
  {D}ata.
\newblock {\em AAAI}, 2017.

\bibitem[\protect\citeauthoryear{Sidana \bgroup \em et al.\egroup
  }{2021}]{2021}
Sumit Sidana, Mikhail Trofimov, Oleh Horodnytskyi, Charlotte Laclau, Yury
  Maximov, and Massih-Reza Amini.
\newblock User preference and embedding learning with implicit feedback for
  recommender systems.
\newblock {\em Data Mining and Knowledge Discovery}, 2021.

\bibitem[\protect\citeauthoryear{Smith and Linden}{2017}]{7927889}
Brent Smith and Greg Linden.
\newblock Two {D}ecades of {R}ecommender {S}ystems at {A}mazon.com.
\newblock {\em IEEE Internet Computing}, 2017.

\bibitem[\protect\citeauthoryear{Wang \bgroup \em et al.\egroup }{2017}]{IRGAN}
Jun Wang, Lantao Yu, Weinan Zhang, Yu~Gong, Yinghui Xu, Benyou Wang, Zhang
  Peng, and Dell Zhang.
\newblock I{RGAN}: {A} {M}inimax {G}ame for {U}nifying {G}enerative and
  {D}iscriminative {I}nformation {R}etrieval {M}odels.
\newblock {\em SIGIR}, 2017.

\bibitem[\protect\citeauthoryear{Wang \bgroup \em et al.\egroup
  }{2018}]{GraphGAN}
Hongwei Wang, Jia Wang, Jialin Wang, Miao Zhao, Weinan Zhang, Fuzheng Zhang,
  Xing Xie, and Minyi Guo.
\newblock Graph{GAN}: {G}raph {R}epresentation {L}earning with {G}enerative
  {A}dversarial {N}ets.
\newblock {\em {AAAI}}, 2018.

\bibitem[\protect\citeauthoryear{Wu \bgroup \em et al.\egroup
  }{2016}]{10.1145/2835776.2835837}
Yao Wu, Christopher DuBois, Alice~X. Zheng, and Martin Ester.
\newblock Collaborative {D}enoising {A}uto-{E}ncoders for {T}op-{N}
  {R}ecommender {S}ystems.
\newblock In {\em Proc. of the 9th ACM Int. Conf. on Web Search and Data
  Mining}, WSDM '16, 2016.

\bibitem[\protect\citeauthoryear{Yu \bgroup \em et al.\egroup }{2019}]{VAEGAN}
Xianwen Yu, Xiaoning Zhang, Yang Cao, and Min Xia.
\newblock V{AEGAN}: {A} {C}ollaborative {F}iltering {F}ramework based on
  {A}dversarial {V}ariational {A}utoencoders.
\newblock {\em IJCAI}, 2019.

\end{thebibliography}
\end{document}